\documentclass[aps,prl,twocolumn,groupedaddress,showpacs]{revtex4}

\begin{document}


\title{Scaling of Spinodal Turbulence between Viscous and Inertial Hydrodynamic Regimes}



\author{R. V. R. Pandya}
\email[]{rvrpturb@uprm.edu}
\affiliation{Department of Mechanical Engineering, University of Puerto
Rico at Mayaguez, Mayaguez,
PR 00680, USA }
\date{\today}

\begin{abstract}
The existence of unique scaling in a crossover regime between viscous and inertial hydrodynamic regimes is revealed for homogeneous, isotropic, incompressible, spinodal turbulence which is characterized, to begin with, by three different length scales and a velocity scale. The obtained scaling exponents are found to be in agreement and in consistency with available simulation results for a broad range of crossover regime. Also, it is observed that the spinodal turbulence in the crossover regime is in complete consistency with the universality class of self-preservation of decaying grid turbulence. We then obtain analytical forms  for various scalings, valid in the crossover regime, through the analysis for self-preservation of spinodal turbulence. 
\end{abstract}


\maketitle

\section{}
A binary fluid homogeneous mixture, when quenched to thermodynamically unstable state, starts separating with growing size of domains of two different components of the binary fluid. Here our focus is on incompressible 50/50 mixture of two fluids having identical viscosity, density and undergoing the separation process. Through experimental, theoretical and simulation work, a general understanding has evolved on various scaling phenomena which unfold during the course of the separation or spinodal decomposition process (see refs. \cite{Furukawa85b,Bray94,HKTF06} and references cited therein). These scaling phenomena can be categorized into four different regimes, namely, 1) pure diffusive, 2) viscous hydrodynamic, 3) inertial hydrodynamic, and 4) crossover regime between viscous and inertial hydrodynamic. Theoretically predicted scalings $L(t) \sim t^{1/3}$ and $L(t) \sim t$ of growing domain size $L$ in time $t$ during the early stage of separation in diffusive and viscous hydrodynamic regimes  \cite{Siggia79}, respectively, are well supported (see refs. \cite{KDBC99,KCPDB01,GNC03} and references cited therein). The pioneer scaling $L(t) \sim t^{2/3}$ in the later stage of inertial hydrodynamic regime \cite{Furukawa85}, though has found support \cite{KDBC99,KCPDB01}, still remains controversial due to the physical requirement of Reynolds number saturation in the long time limit and which has led to exponent values other than $2/3$ \cite{GE99,SC00,SG05}. Contrary to this, we also note that scaling theory of Kendon \cite{Kendon00} obtains $2/3$ and is consistent with Reynolds number saturation. Further, a broad range of crossover between viscous and inertial hydrodynamic regimes exists as exhibited in simulation studies \cite{KDBC99,KCPDB01}. It is clearly apparent from the  results of these simulation studies that scaling does exists in the crossover regime, but so far it has not been dealt with and predicted theoretically. In this letter, we reveal existence of unique scaling $L \sim t^{3/4}$ in the broad crossover regime, connecting viscous and inertial hydrodynamic regimes where viscous and inertial effects are both important. This scaling along with other obtained scalings for velocity and its length scales, Taylor-scale and Kolmogorov dissipation scale are found to be in agreement with the simulation data of Kendon et al. \cite{KCPDB01} for spinodal turbulence and in complete consistency with results for decaying of homogeneous, isotropic turbulence \cite{Batchelor67}, as well.  For our analysis, we use theoretical framework of Kendon \cite{Kendon00}.   

The Navier-Stokes equation governing the phenomena of spinodal turbulence velocity field ${\bf v}$ in incompressible, isothermal 50/50 mixture of two fluids of density $\rho$ can be written as
\begin{equation}
\rho\frac{\partial {\bf v}}{\partial t}+\rho({\bf v}\cdot {\bf \nabla}){\bf v}=-{\bf \nabla}p+\eta \nabla^2{\bf v}-{\bf \nabla}{\cal P}. \label{nse}
\end{equation}
Here $p$ is pressure, $\eta$ is viscosity and $-{\bf \nabla}{\cal P}$ is the interfacial force density, usually approximated by $\sigma/L^2$ \cite{Furukawa85,Siggia79} where $\sigma$ is the interfacial tension between the two fluids and $L$ is length scale representing the domain size. For this homogeneous isotropic case, the kinetic energy of turbulence per unit volume, i.e. $\frac{\rho}{2}\langle {\bf v}\cdot {\bf v}\rangle $, is governed by
\begin{equation}
\frac{d}{dt}\rho\langle {\bf v}\cdot {\bf v}\rangle/2=-\eta \langle({\bf \nabla v})^2 \rangle +\langle -{\bf v}\cdot {\bf \nabla}{\cal P} \rangle \label{ke}
\end{equation} 
where $\langle \, \rangle$ denotes ensemble average, the first term on the right hand side (rhs) represents viscous dissipation rate and the second source term on the rhs accounts for the energy transfer rate to the fluid arising from the interfacial force. This source term is approximated, so far, by $\sigma \dot{L}/L^2$ \cite{Kendon00} for the scaling purpose and implicitly assumes the approximation ${\bf v}\sim\dot{L}\sim L/t$. We should mention that we do not use such an approximation for ${\bf v}$ when a separate scaling (see scaling given by (\ref{s2})) is used for it. In this respect we allow Navier-Stokes equation, governing the phenomena, to govern and yield length scale $L_{v}\sim |{\bf v}|t$ (or ${\bf v}\sim \dot{L_v}$) of the velocity.  

Now consider the scaling behavior of various quantities, namely, three length scales $L, L_\nabla, L_{\nabla^2}$ and velocity scale ${\bf v}$, as
\begin{eqnarray}
{\mbox {domain size:}}   &&L\sim t^\alpha, \label{s1} \\
{\mbox {fluid velocity:}}   && {\bf v}\sim t^{\beta},  \label{s2}\\
{\mbox {velocity first derivative:}}   &&{\bf \nabla}{\bf v} \sim\frac{{\bf v}}{L_\nabla}\sim t^{\beta-\alpha'}, \\
{\mbox {velocity second derivative:}}  &&{\nabla^2}{\bf v} \sim\frac{{\bf v}}{L_{\nabla^2}^2}\sim t^{\beta-2\alpha''}.\label{s4}  
\end{eqnarray}
Further, using scalings $L\sim t^\alpha$, ${\bf v}\sim t^{\beta}$ and approximation $-{\bf \nabla}{\cal P} \sim \sigma/L^2$, we can write for the source term
\begin{equation}
\langle -{\bf v}\cdot {\bf \nabla}{\cal P} \rangle \sim \sigma \frac{t^\beta}{L^2}\sim \sigma t^{\beta-2\alpha}\label{source}
\end{equation}
and which is different than the usually employed approximation $\sigma \dot{L}/L^2 \sim \sigma t^{-\alpha-1}$ by Kendon \cite{Kendon00}. We should mention that this new relation (\ref{source}) leads  us to a unique set of values for all scaling exponents ($\alpha, \beta, \alpha', \alpha''$) and yields a situation where scaling exponents of all terms in the Navier-Stokes equation (\ref{nse}) are identical. For this reason, we have associated the obtained unique scaling to the crossover regime, between viscous and inertial hydrodynamic regimes, in which all terms in the Navier-Stokes equation are relevant to the involved scaling phenomenon.

Using scaling behaviors given by (\ref{s1})-(\ref{source}) allow us to write the turbulent kinetic energy Eq. (\ref{ke}) in terms of power of $t$ as
\begin{equation}
\rho \beta t^{2\beta-1} \sim -\eta t^{2\beta -2\alpha'}+\sigma t^{\beta-2\alpha}
\end{equation}
and which further suggests
\begin{equation}
{2\beta-1}={2\beta -2\alpha'}={\beta-2\alpha}. 
\end{equation}
This gives 
\begin{equation}
\alpha'=1/2, \quad \beta=1-2\alpha. \label{value}
\end{equation}

Now we study the behavior of Navier-Stokes equation (\ref{nse}) when various scalings (\ref{s1})-(\ref{s4}) are substituted in it. It should be noted that in the absence of $-{\bf \nabla}{\cal P}$, scaling of $-{\bf \nabla}p$ is identical to the scaling of convection term, i.e., second term on the left hand side of Eq. (\ref{nse}). We make use of this fact along with (\ref{s1})-(\ref{s4}) and approximation $-{\bf \nabla}{\cal P} \sim \sigma/L^2$ to obtain, from the Navier-Stokes equation,
\begin{equation}
\rho \beta t^{\beta-1}+\rho t^{2\beta-\alpha'}\sim \eta t^{\beta-2\alpha''}+\sigma t^{-2\alpha}. \label{scns}
\end{equation}
Further making use of (\ref{value}) into (\ref{scns}), we obtain
\begin{equation}
\rho \beta t^{-2\alpha}+\rho t^{\frac{3}{2}-4\alpha}\sim \eta t^{1-2\alpha-2\alpha''}+\sigma t^{-2\alpha}
\end{equation}
which suggests
\begin{equation}
-2\alpha= -4\alpha+3/2=1-2\alpha-2\alpha''
\end{equation}
and yields a unique solution 
\begin{equation}
\alpha = 3/4, \quad \alpha''=1/2.
\end{equation}
Subsequently from (\ref{value}) we obtain 
\begin{equation}
\beta=-1/2. \label{beta}
\end{equation} 
This unique scaling solution suggests that all the terms, including pressure term, in the Navier-Stokes equation are scaled identically as $t^{-3/2}$. For the purpose of completeness, we must mention that the obtained unique set of scaling exponents ($\alpha=3/4,\beta=-1/2,\alpha'=1/2,\alpha''=1/2$) along with $ -{\bf v}\cdot {\bf \nabla}{\cal P} \sim \sigma \frac{t^\beta}{L^2}\sim \sigma t^{\beta-2\alpha}$ also provide identical scaling exponent for all the terms in the equation for local energy \cite {Kendon00} or instantaneous kinetic energy of turbulence $\rho {\bf v}\cdot {\bf v}/2$. This further strengthen the association of obtained scalings to the crossover regime in which all the terms in the equation for local energy become crucially important.

A few more scaling relations for velocity length scale $L_v$, two different Reynolds numbers ($Re_L$, $Re_{\bf v}$), dissipation rate $\epsilon$, Taylor-scale $\lambda$ and Kolmogorov dissipation scale $\lambda_d$ can be derived using the above scaling exponents. These are
\begin{eqnarray}
L_v &\sim & |{\bf v}|t \sim t^{1/2} \\
Re_L&=&\frac{\rho|{\bf v}|L}{\eta} \sim  t^{1/4},\\
Re_{\bf v}&=&\frac{\rho|({\bf v}\cdot {\bf \nabla}){\bf v}|}{\eta |\nabla^2{\bf v}|} \sim  t^0,\\
\epsilon &=&\eta\langle({\bf \nabla}{\bf v})^2 \rangle  \sim  t^{-2}, \label{diss}\\
\lambda &=&\left(\frac{5\eta\langle |{\bf v}|^2\rangle}{\epsilon}\right)^{1/2}\sim  {L_\nabla} \sim  t^{1/2}, \label{lam}\\
\lambda_d&=&\left(\frac{\eta^3}{\rho^2 \epsilon}\right)^{1/4}  \sim   t^{1/2}. \label{lamd}
\end{eqnarray}
It should be noted that scaling for velocity length scale $L_v \sim t^{1/2}$ turns out to be different than the scaling for domain size $L \sim t^{3/4}$. In fact, this prediction is consistent with simulation study by Kendon et al. \cite{KCPDB01} as exhibited in their figure (11a) in the crossover regime $10^2< t < 10^6$. Thus in view of this, the traditional approximation $\sigma\dot{L}/L^2$ for $\langle -{\bf v}\cdot {\bf \nabla}{\cal P} \rangle$ which employs $|{\bf v}|\sim \dot{L}$ is not valid in the crossover regime. And our choice of approximation as given by (\ref{source}) turns out to be correct for the crossover regime.  

Now important question about the validness of various obtained scalings arises. The available simulation study \cite{KCPDB01} (hereafter referred to as KCPDB paper) provides information for a long range of crossover regime ($10^2 < t <10^6$ or $1 < Re_L < 50$ ). In KCPDB, results for $L$ as shown in figure 6 and values of $\alpha$ listed in Table 5 suggest that our predicted exponent $\alpha=3/4$ is a very reasonable value in the crossover region. Result of figure (17a) in KCPDB provide scaling close to $t^{-2}$ for dissipation rate and is in agreement with present scaling given by (\ref{diss}). It is encouraging to note that the predicted scaling exponent $1/2$, given above in (\ref{lam}) and (\ref{lamd}), for Taylor-scale and Kolmogorov dissipation scale are in agreement with crossover regime data of figure (18b) in KCPDB. Further, the present scaling prediction $t^{-3/2}$ for each term of Navier-Stokes equation appears visually to be consistent with data of figure (18a) in KCPDB.

Apart from the consistency with spinodal turbulence simulation data, we note that the obtained scalings are also consistent with decaying, homogeneous isotropic turbulence. In this case, when the turbulence velocity scale is decaying as ${\bf v} \sim t^{\beta}$ with $\beta$ as a constant, $\lambda \sim t^{1/2}$ and integral length scale of velocity $\sim t^{1/2}$ (see e.g. refs. \cite{Batchelor67,WG02}). So these exponents for $\lambda$  and the integral length scale of velocity are exactly the same as predicted above if we associate $L_v$ to the integral length scale. In addition to this, the above predicted scaling exponent for velocity decay $\beta=-1/2$, scalings of dissipation rate $\sim t^{-2}$ and Kolmogorov dissipation scale $\sim t^{1/2}$ are identical to those belonging to universality class of {\it initial period} of decay of grid turbulence preserving the shape of whole of the energy spectrum and correlations functions, except for the portion at very smallest values for magnitude of wave numbers \cite{Batchelor67}. In view of this consistency with the universality class, we can conclude that spinodal turbulence is self-preserving in the cross over regime. Now we provide analysis on self-preservation of spinodal turbulence which obtains exact analytical forms for various scalings consistent with the scaling exponents obtained above. 

We should note that though we started with three length scales, the scaling exponents for $L_\nabla$ and $L_{\nabla^2}$ are identical. This suggests reduction by one in number of length scales to characterize crossover regime of spinodal turbulence. For the analysis of self-preservation, we consider equation for
\begin{equation}  
R_{ii}({\bf r},t)=\langle {\bf v}({\bf x},t)\cdot {\bf v}({\bf x}+{\bf r},t)\rangle =u^2\left(3f(r,t)+r\frac{\partial f}{\partial r}\right)
\end{equation}
for homogeneous isotropic turbulence which can be obtained from Eq. (\ref{nse}) \cite{Batchelor67} and can be written as
\begin{eqnarray}
\rho \frac{\partial R_{ii}}{\partial t}=\rho u^3\left(r\frac{\partial }{\partial r}+3\right)\left(\frac{\partial}{\partial r}+\frac{4}{r}\right)k(r,t)+\nonumber \\
+2\eta \left(\frac{\partial^2}{\partial r^2}+\frac{2}{r}\frac{\partial }{\partial r}\right)R_{ii} +2 \langle {\bf v}({\bf x},t) \cdot {\bf s}({\bf x}+{\bf r},t)\rangle. \label{sp01}
\end{eqnarray}
Here $u^2=\langle {\bf v}\cdot {\bf v}\rangle/3$, $R_{ij}$ is two-point velocity correlation, $r=|{\bf r}|$ is distance between two points for two-point correlation, $f$ is the longitudinal velocity correlation coefficient, $k(r,t)$ is single scalar function determining the triple-velocity correlations (see ref. \cite{Batchelor67}) and  ${\bf s}=-{\bf \nabla}{\cal P}$ represents interfacial force per unit volume. Also, for the homogeneous isotropic turbulence we can write $\langle {\bf v}({\bf x},t) \cdot {\bf s}({\bf x}+{\bf r},t)\rangle = u s {\cal F}(r,t)$ where $s^2=\langle {\bf s}\cdot {\bf s}\rangle /3$.

Now we consider characteristic length scale $l\equiv l(t)$ (which will turn out to be similar to $L_{\nabla}$), introduce a dimensionless variable $\psi=r/l$ and take self-preserving form for functions $f$, $k$, ${\cal F}$ as
\begin{equation}
f\equiv f(\psi);\,\, k\equiv k(\psi);\,\, {\cal F}\equiv {\cal F}(\psi).
\end{equation}
Substituting these in Eq. (\ref{sp01}), we obtain
\begin{eqnarray}
\rho \frac{du^2}{dt} F_1(\psi) +\frac{\rho u^2}{l}\frac{dl}{dt} F_2(\psi)&=&\frac{\rho u^3}{l} F_3(\psi)  +\frac{\eta u^2}{l^2}F_4(\psi)\nonumber \\&+&u s {\cal F}(\psi) \label{sp02}
\end{eqnarray}
where $F_1,F_2,F_3$ and $F_4$ are some functions of $\psi$ only and their details are not presented here as they are not relevant for further analysis. For self-preservation, Eq. (\ref{sp02}) suggests that all ratios between different factors multiplying $F_1,F_2,F_3, F_4$ and ${\cal F}$ should be constants. For obtaining analytical solutions,  consider  
\begin{eqnarray}
\frac{\rho u^2}{l}\frac{dl}{dt} &=&  A_1 \frac{\eta u^2}{l^2}, \label{spp1}\\
\frac{\rho u^3}{l} &=& A_2 \frac{\eta u^2}{l^2}, \label{spp2}\\
\rho \frac{du^2}{dt} &=& A_3 \frac{\rho u^2}{l}\frac{dl}{dt}, \label{spp3} \\
u s&=& A_4 \frac{\rho u^3}{l} \label{spp4}
\end{eqnarray}  
where $A_1,A_2,A_3$ and $A_4$ are constants. These Eqs. (\ref{spp1})-(\ref{spp4}) suggest analytical solutions, written as
\begin{equation}
l=\left[\frac{2 A_1\eta}{\rho}(t-t_0)+l_0^2\right]^{1/2},
\end{equation}
\begin{equation}
u=\frac{A_2\eta}{\rho}\left[\frac{ 2A_1\eta}{\rho}(t-t_0)+l_0^2\right]^{-1/2},
\end{equation}
\begin{equation}
A_3=-1/2,
\end{equation}
\begin{equation}
s=\frac{A_4}{\rho}(A_2\eta)^2\left[\frac{2 A_1\eta}{\rho}(t-t_0)+l_0^2\right]^{-3/2}.
\end{equation}
Here $l_0$ is value of length scale $l$ at time $t_0$ which is in the crossover regime. It should be noted that $l \sim t^{1/2}$ and thus is similar to $L_\nabla$. Also, $s\sim t^{-3/2}$ is consistent with the scaling $L\sim t^{3/4}$ as $s\sim \sigma/L^2$.

In conclusion, we have suggested existence and values of the unique scaling exponents in the crossover regime for the phenomena of spinodal turbulence where all forces in the Navier-Stokes equation are important and scale to identical exponent. The obtained scalings have been found to be in reasonable agreement with the simulation data and is in complete consistency with the self-preserving phenomena of decay of grid turbulence during the initial period. Further, analytical forms for characteristic length, velocity scales ($l$ and $u$, respectively) and rms of interfacial force density component, i.e. $s$, are obtained through the analysis for self-preservation of spinodal turbulence in the crossover regime. 
Though there exists a few studies on the crossover regime \cite{KCPDB01,PWC02}, it is hoped that the present work would stimulate further more focused studies on the scaling behaviour of crossover regime of the spinodal turbulence so as to rigorously verify or dispute the scaling exponents predicted theoretically in this letter.        
\begin{acknowledgments}
I am very much thankful to my friend Dr. Paul Stansell for introducing me to this subject of binary fluid demixing.  
\end{acknowledgments}


\end{document}